\documentclass[12pt]{article}
\usepackage{amsmath,amssymb,latexsym,theorem,bbm,bm}
\def\'{\futurelet\next\accuteacc}
\def\accuteacc#1{\ifx i\next\accent19\char16\relax\else\accent19\next\fi}
\let\~=\H
\let\==\H

\def\veps{\varepsilon}

\newcommand{\beq}{\begin{equation}}
\newcommand{\eeq}{\end{equation}}
\newcommand{\beqa}{\begin{eqnarray}}
\newcommand{\eeqa}{\end{eqnarray}}

\title{
Testing goodness-of-fit of random graph models}

\author{
Vill\H{o} Csisz\'ar$^1$
\and
P\'eter Hussami$^2$
\and
J\'anos Koml\'os$^3$
\and
Tam\'as F. M\'ori$^{1,5}$
\and
L\'idia Rejt\H{o}$^{2,4}$
\and
G\'abor Tusn\'ady$^2$
}

\date{submitted: May 4, 2012, \\revised: November 8. 2012}

\begin{document}
\maketitle

\begin{abstract}
Random graphs are matrices with independent $0-1$
elements with probabilities determined by a small number
of parameters. One of the oldest model is the Rasch model
where the odds are ratios of positive numbers scaling
the rows and columns. Later Persi Diaconis with his coworkers
rediscovered the model for symmetric matrices and called
the model beta. Here we give goodnes-of-fit tests for
the model and extend the model to a version of the block
model introduced by Holland, Laskey, and Leinhard.  

\end{abstract}

\section{{\bf{Introduction }}}

\footnotetext[1]{E\"otv\"os Lor\'and University, Budapest, Hungary}
\footnotetext[2]{Alfr\'ed  R\'enyi
Mathematical Institute of the Hungarian Academy of Sciences,
Budapest,Hungary}
\footnotetext[3]{Rutgers University, Department of Mathematics,
New Brunswick, New Jersey, USA}
\footnotetext[4]{University of Delaware, Statistics
Program, FREC, CANR, Newark, Delaware, USA}
\footnotetext[5]{Tam\'as F. M\'ori's research was supported by OTKA
grant 12574}

\bigskip

Let $n$ be a positive integer, $1\leq i,j\leq n$, and $\veps(i,j)$
independent
random variables such that $\veps(i,j)=\veps(j,i)$ and $\veps(i,i)=0$,
furthermore
\beq \label{add}
P(\veps(i,j)=1) = p_{i,j} = p + p_i + p_j, \quad 1\leq i<j \leq n,
\eeq
where the sum of the $p_i$-s is zero. The least square estimate 
$\hat p$ of $p$ is
the average of the epsilons, and the least square estimate of $p_i$ is
the average of the differences $\veps(i,j)-\hat p$. The modification
of the model for non-symmetric matrices is straightforward, and in
that case the statistical inference is practically a two-way
analysis of variance. Perhaps this is the simplest random graph 
model but it shares the inconvenient property of many other random graph
models that it is hard to ensure that edge probabilities remain
in the interval $(0,1)$. If we use the odds
\beq \label{odds}
r_{i,j} = {\frac {p_{i,j}}{1-p_{i,j}}},
\eeq
instead of the probabilities,
then it is enough to ensure the positivity of $r_{i,j}$-s. This 
is the case in
the model introduced by George Rasch {\cite {R}}. Historically
the odds were defined as the ratios of scaling factors for rows
and columns but we prefer the multiplicative form
\beq \label{bg}
r_{i,j} = \beta_i \gamma_j
\eeq
for non-symmetric and
\beq \label{bet}
r_{i,j} = \beta_i \beta_j
\eeq
for symmetric case. Statistical investigation of the model started
with Andersen {\cite {A}} (see also \cite {{L}, {Pon}, {V}})
and later Persi Diaconis with his coworkers
rediscovered the model and introduced the name {\it {beta-model}}
for its parameter. The model has many attractive properties
(see in \cite {{BC}, {BH}, {D8}, {BD}, {CDS}, {OHT}}):

-- degree sequences are sufficient statistics

-- the model covers practically all possible expected degree sequence

-- the conditional distribution of the graphs on condition of a
prescribed degree sequence is uniform on the set of all graphs
with the given degree sequences.

Statistically inference emerged from Gaussian distribution and
later was extended to random variables in Euclidean spaces but
the statistical inference on discrete structures is rather sparse
(\cite {{BT}, {CSRT}, {CHKRMT}, {HUS}, {NNTB}}). Mathematical investigation of
graphs has its own history. Nowadays instead of graphs we are
speaking of networks (\cite {NBW}) where the most investigated
model is the stochastic block model introduced by Holland, Laskey,
and Leinhard ({\cite {SBM}). Here the vertices are labeled by
small numbers or colors and edge probabilities depend only on
the labels (\cite {{BCCZ}, {FP}}). With an eye on preferential attachment where
degree sequences follow scale-free power-law the block model
was criticized because it has moderated flexibility on
degree sequences. Chung, Lu, and Vu {\cite {CLV}} introduced  
a model with independent vertices, Chaughuri, Chung, and Tsiatas
(\cite {CCT}) introduced the {\it planted partition model}
(see also \cite {MNS}).
Karrer and Newman {\cite {KN}} proposed and other extension
of the block model. A natural extension of these models is
the unification of the beta and block models:
\beq \label{bb}
r_{i,j} = b(i,c(j)) b(j,c(i)),
\eeq
where
$b(.,.)$ is a positive matrix with $n$ rows and $k$ columns, and
$c(i)$ is the label of the $i$-th vertex i.e. it is
an integer between $1$ and $k$. We call the model {\it k-beta model.}
The estimation of the labels in block models is possible
by the spectral method (\cite {RCY}). It is generally believed
that eigenvectors and eigenvalues of the matrix $\veps(i,j)$
tells everything of the structure of the graph 
(\cite {{CCT}, {FC}, {FL}, {LL}, {LP}, {NaNe}}),
while there are many attempts to provide more flexible models
(\cite {{ChDia}, {PLV}}).

\section{\bf Goodness-of-fit}

We can not test edge-independence on a single graph. While i.i.d.
sample is common in statistical inference, in case of graphs the
sample generally means a copy of a graph. Perhaps the number one
question in statistical inference is the following. Let
\beq \label{prob}
p_1,\ldots,p_n
\eeq
be an arbitrary given sequence of probabilities, and
\beq \label{eps}
\veps_1,\ldots,\veps_n
\eeq
be independent $0-1$ variables such that $P(\veps_i = 1) = p_i$.
Can we test the model? A randomized answer is the following. Let
\beq \label{unic}
u_1,\ldots,u_n
\eeq
independent and uniformly distributed in $(0,1)$. Then
\beq
x_i = p_i u_i \veps_i + (1-\veps_i)(p_i+(1-p_i) u_i), 
\quad i=1,\ldots,n
\eeq
are independent and uniformly distributed in $(0,1)$, what we
can test. An other, more practical solution is ordering the
the pairs $(p_i,\veps_i)$ according to the $p_i$-s in increasing
order and compare their partial sums. Or we can clump them
into blocks of small number and compare again the sums.
All these possibilities hold for graphs with estimated
edge probabilities. Let us partition the edges of the complete graph
according to the blocks formed with respect to the edge probabilities.
In each portion the edge probabilities are close to each other
whence the $\veps_{i,j}$-s corresponding to that portion
behave like a pure random graph. what we again can test
e.g. by their sums on subsets of vertices.

Blitzstein and Diaconis (\cite {{BD}, {CDHL}}) propose for testing the
beta model the following general procedure. Let us choose
any graph statistic and determine it on our graph. Let us
generate as many graph we can with the same degree sequence
as the investigated graph has according to the uniform
distribution, and let us calculate the chosen statistics.
If the value of the sample graph is inside the generated numbers,
we accept the beta model, otherwise reject it. One can ask,
are there any effect of the choose on the power of the test?

We have found by computer simulations that graphs generated by
beta model have only one eigenvalue proportional with $n$,
all the others are of order $\sqrt n$. We think that it is
a characteristic property of beta graphs. One wonders that

-- if beta model covers all possible degree sequences

-- the conditional distribution is uniform over graphs
sharing the same degree sequence,

\noindent
then how is possible that graph behaves differently from
typical graphs generated by beta model? Of course there
are graphs having many large eigenvalues. But where are
they coming from once beta model can generate all the graphs?
A possible solution of the catch is the following.

Let us generate a meta graph from graphs sharing the same
degree sequence. Let us say that neighborhood in this
meta graph is given by on single swap. If we have four
vertices A, B, C, D in a graph such that AC, BD is and
edge but AD, BC is not, then changing existence into non
existence among these edges we form a new graph with the same
degree sequence. The degree of a graph in this meta graph
goes parallel with the second largest eigenvalue: typical
beta model graphs have minimal degree and any increase in
their degree results in a more complicated eigenvalue structure.
Perhaps the degree in the meta graph is the most characteristic
statistic for beta model.

\section{\bf The k-beta model}

The maximum likelihood equations for the parameters $b(.,.)$
in (5) say that the expected values of degrees
{\it inside} all the subgraph with a given pair of labels
should be the same us in the given graph. This is the case
when the labels are known. With unknown labels we can form
a two-level optimization: for each label set first to determine
the parameters $b(.,.)$ next changing a small number of labels
and repeat the calculation of the parameters. But the procedure
is slow even for graphs of moderate sizes. Spectral methods
available for block models fail for coloring k-beta models because
the model lose the well pronounced checkerboard character
of block models. It is the ANOVA what offers an applicable
algorithm. For any set $C$ of labels $c(.)$ let us calculate
the statistic
\beq \label{ANOVA}
Q(C) =
\sum^n_{i=2}\sum^{i-1}_{j=1} (\veps(i,j) - u(c(i),c(j)) - v(i,c(j))
- v(j,c(i)))^2,
\eeq
where
\beq \label{AVE}
u(s,t) = {\frac {\sum_{c(i)=s}\sum_{c(j)=t} \veps(i,j)}
                {\sum_{c(i)=s}\sum_{c(j)=t} 1}},
\eeq
and
\beq \label{PAVE}
v(i,t) = {\frac {\sum_{c(j)=t} (\veps(i,j)-u((c(i),t))}
                {\sum_{c(j)=t} 1}}.
\eeq
$Q(C)$ is the sum of two way ANOVA sum of squares calculated
independently for subgraphs defined for pairs of labels. Starting
from a uniform random set $C$ of labels on the vertices and
perturbing small number of labels in the individual steps
a simple greedy optimization results in a good set of labels,
which is close to the original (true) labels.

For evaluating the character of a random graph we use the number
\beq \label{ent}
\exp(-{\frac {\sum^n_{i=2}\sum^{i-1}_{j=1} (p(i,j)\log p(i,j) +
(1-p(i,j))\log (1-p(i,j))}{n(n-1)/2}})
\eeq 
We call it {\it delogarithmed average entropy} or DAE.
This is a number between $1$ and $2$. If it is close to one
the graph is almost deterministic: the probabilities
are close to $0$ or $1$. In checkerboard block models
it means that empty and full subgraphs are amalgamated together.
If DAE is close to $2$ then the graph has no structure at al.
DAE depends on edge density, too. The above tendency is
valid for edge density ${\frac {1}{2}}$, for other edge
densities the cut point is closer to $1$. According to our
experience if DAE is smaller then $1.9$ while edge density 
is half, then we are able to reconstruct the original labels.
For these graphs the number of non-trivial eigenvalues is
$2k-1$, thus the spectrum determines the number of different
labels.

The k-beta model has a sister model
\beq \label{RANK}
r_{i,j} = \sum^k_{s=1} b(i,s) b(j,s)
\eeq
what we call {\it small odds rank} model. Strictly speaking
we ought to redefine the diagonal of odds matrix, but perhaps
the name is permissible without doing so. The 
maximum likelihood estimation of parameters
in small odds rank models is straightforward and
the block structure is detectable in the estimated parameters.
Actually the block model is in the intersection of k-beta
and small odds rank models, thus if there is any block
structure in the graph it is detectable even in fitting
k-beta model to the graph. But if there is no block structure
and we are trying to use ANOVA coloring for a small odds rank
graph then the algorithm is no longer stable, it results in
different local minima in each runs.

\end{document}